\documentstyle[11pt,paspconf,epsf]{article}
\begin{document}

\title{Magnetic Fields on the Dynamics of the ICM}
\author{Denise R. Gon\c calves \thanks {denise@ll.ia.es}} 
\affil{Instituto de Astrof\'\i sica de Canarias, E-38200 La Laguna, 
Tenerife, Spain}

\author{Am\^ancio C. S. Fria\c ca}
\affil{Instituto Astron\^omico e Geof\'\i sico - USP, Av. Miguel Stefano
4200, 04301-904 S\~ao Paulo - SP, Brazil}

\begin{abstract}
Could the discrepancies found in the determination of mass in clusters of
galaxies, from gravitational lensing data and from X-rays observations, be
consequence of the standard description of the ICM, in which it is assumed
hydrostatic equilibrium maintained by thermal pressure? In analogy to the
interstellar medium of the Galaxy, it is expected a non-thermal term of
pressure, which contains contributions of magnetic fields. We follow the
evolution of the ICM, considering a term of magnetic pressure, aiming at
answering the question whether or not these discrepancies can be explained
via non-thermal terms of pressure. Our results suggest that the magnetic
pressure could only affect the dynamics of the ICM on scales as small as 
$\la 1 {\rm kpc}$. These results are compared to the observations of large
and small scale magnetic fields and we are successful at reproducing
the available data.
\end{abstract}

\section{Introduction}

Since the work of Loeb \& Mao (1994), the possibility of
explaining the discrepancies on mass determinations, found by
Miralda-Escud\'{e} \& Babul (1994), via non-thermal pressure support has
been widely discussed. The discrepancy arises
from the two most promising techniques to obtain clusters of galaxies
masses. On one hand, the determination of masses in clusters of galaxies,
via X-ray data, is based on the hypothesis that the ICM is in hydrostatic
equilibrium with the gravitational potential, using the radial profiles of
density and temperature (Nulsen \& B\"{o}hringer
1995). On the other hand, gravitational lensing measures the projected
surface density of matter, a method which makes no assumptions on the
dynamical state of the gravitating matter 
(Miralda-Escud\'{e} \& Babul 1994; Smail et al. 1997). 

In clusters with diffuse radio emission X-ray observations can give a lower
limit to the strength of the magnetic field (the $3{\rm K}$ background photons 
scattering off the relativistic electrons produces the diffuse X-ray 
emission). Typically, this limit is $B\geq
0.1\;{\rm \mu G}$ (Rephaeli et al. 1987) on scales of $\sim 1\;{\rm Mpc}$.
Such a kind of detection of clusters magnetic fields leads,
using ROSAT\ PSPC data and also ${327}{\rm MHz}$ radio map of Abell 85, to
an estimate of $(0.95\pm .10)\;{\rm \mu G}$ (Bagchi et al. 1998). 
In the case of Faraday rotation the information obtained is the upper limit
on the intensity of the field, and the measured values are $(RM\leq 100\;%
{\rm rad/m^2})$, which is more or less consistent with a intracluster field
of $B\sim 1\;{\rm \mu G}$, with a coherence length of $l_B\leq 10\;{\rm kpc}$.
This strength of the magnetic field corresponds to a ratio of magnetic to
gas pressure $p_B/p_{gas}$ $\leq 10^{-3}$, implying that $B$ does not
influence the cluster dynamics (at least on large scales). In inner regions
the magnetic fields are expected to be amplified due to the gas compression
(Soker \& Sarazin 1990). For frozen-in fields and homogeneous and spherically 
symmetric inflow, $B\propto r^{-1}$ and $RM\propto
r^{-1}$, ($p_B\propto r^{-2}$ whereas gas pressure increases).
Very strong Faraday rotations were observed ($RM\sim 4000\;{\rm rad/m^2}$)
implying $B\geq 10\;{\rm \mu G}$ at $l_B\sim 1\;{\rm kpc}$ (Taylor \& Perley
1993; Ge \& Owen 1993, 1994).

\section{Evolution of the ICM with Magnetic Pressure}

Using a spherically symmetric finite-difference scheme Eulerian code, 
the evolution of the intracluster 
gas is obtained by solving the
hydrodynamic equations of mass, momentum and energy conservation (see Fria\c ca
1993),  coupled to the state equation for a fully ionized gas with $10\%$
helium by number. The mass distribution, $M(r)$, is due to the contribution
of the X-rays emitting gas plus the cluster collisionless matter (which is
the sum of the contributions of galaxies and dark matter -- the latter being
dominant) following $\rho_{cl}(r)=\rho _c(1+r^2/a^2)^{-3/2}$,  
{$\rho _c$ and $a$ (the cluster
core radius) are related to the line-of-sight velocity dispersion, $\sigma $,
 by $9\sigma
^2=4\pi Ga^2\rho _c$. }The total pressure $p_t$ is the sum of thermal and
magnetic pressure, e.g. $p_t=p+p_B$. The constraints to the magnetic
pressure come from observations, from which $p_B=B^2/8\pi \simeq 4\times
10^{-14}{\rm erg\;cm}^{-3}{\rm s}^{-}1$ (cf. Bagchi et al. 1998) for a diffuse
field located at $\sim 700h_{50}^{-1}{\rm kpc}$ from the cluster center.

The initial conditions for the gas are an isothermal atmosphere ({$%
T_0=10^7$}${\rm K}$) with $30\%$ solar abundance and density distribution
following that of the cluster dark matter. The evolution is followed until
the age of $14{\rm \;Gyr}$. 
We assume: frozen-in field; spherical symmetry for the flow and the cluster
itself; and that at $r>r_c$ (the cooling radius), the magnetic
field is isotropic, i.e., $B_{r^{}}^2=B_b^2/2=B^2/3$ and $l_r=l_t\equiv l$
(where $B_r$ and $B_t$ are the radial and transversal components of the
magnetic field $B$ and $l_r$ and $l_t$ are the coherence length of the
large-scale field in the radial and transverse directions). In order to
calculate $B_r$ and $B_t$ for $r<r_c$ we modified the calculation of the
magnetic field of Soker \& Sarazin (1990) by considering an inhomogeneous
cooling flow (i.e. $\dot{M}_i\neq \dot{M}$ varies with $r$). Therefore, the
two components of the field are then given by $D/Dt(B_{r^{}}^2r^4\dot{M}%
^{-2})=0$ and $D/Dt(B_{t^{}}^2r^2u^2\dot{M}^{-1})=0$. In our models it is
admitted that the reference radius is the cooling radius $r_c$. In fact, we
modify the geometry of the field when and where the cooling time comes to be
less than $10^{10}{\rm yr}$. Therefore, our condition to assume a
non-isotropic field is $t_{coo}\equiv 3k_BT/2\mu m_H\Lambda (T)\rho \ga %
10^{10}{\rm yr}$.

\section{Models and Results}

There are four
parameters to consider in each one of the models: $\sigma =1000{\rm \;km/s}$,
the cluster velocity dispersion; $\rho _0${{$=1.5\times 10^{-28}$}}${\rm g\;cm}%
^{-3}$, the initial average mass density of the gas; $a=250\;{\rm kpc}$, the
cluster core radius; and $\beta _0=10^{-2}$ (model A)$,10^{-3}$ (model B),
the initial magnetic to thermal pressure ratio. 

First of all, the evolution we follow here is characteristic of cooling
flow clusters and in this scenario we discuss the evolution of the basic
thermodynamics parameters. Considering the overall characteristics of our
models, we compare the present models with Peres et al. (1998) deprojection
results (based on ROSAT observations), pointing out that the central cooling 
time here adopted as our cooling
flow criterion, e.g. $t_{coo}\ga 10^{10}{\rm yr}$, is typical for a fraction
between $70\%$ and $90\%$ of their sample. This allow us conclude that our
models, which present cooling flows since the cluster has the age of $\sim
7-9{\rm \;Gyr}$, are typical for their sample.

\begin{figure}
\plottwo{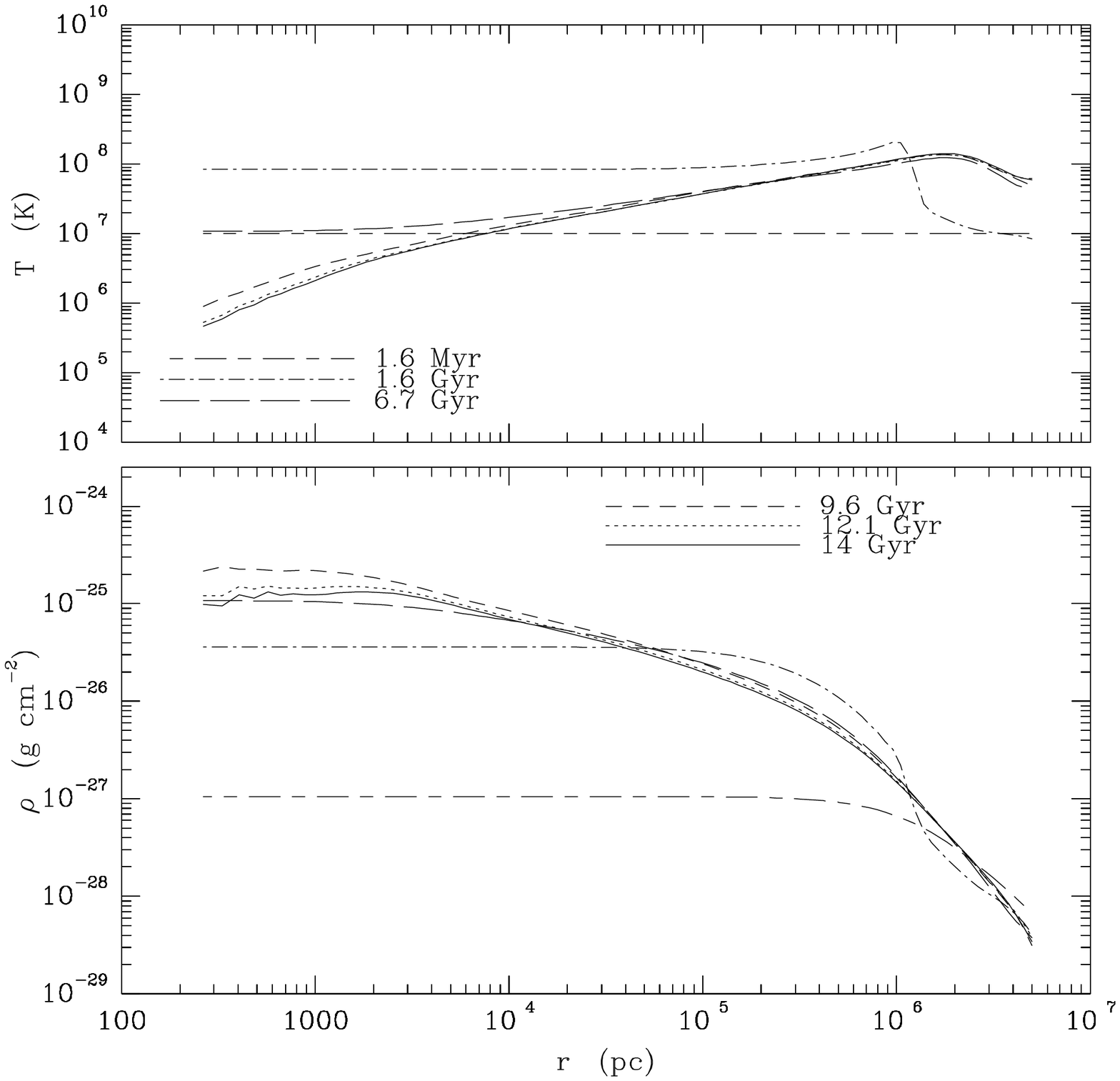}{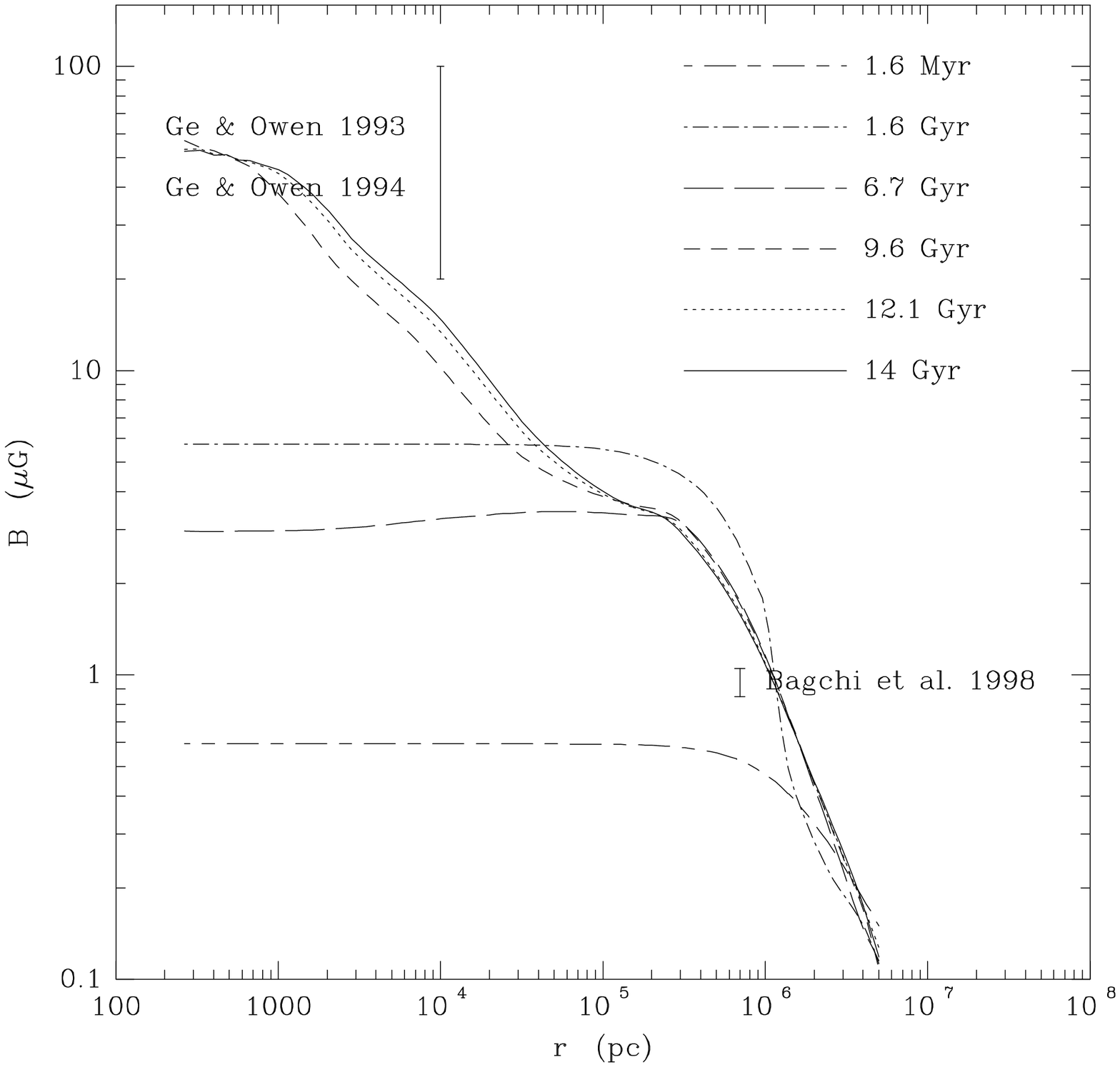}
\caption{Evolution of the density and temperature (left) and magnetic field 
strength (right) profiles. Curves represent
early and late stages of the ICM evolution, as labeled, for the model A.}
\end{figure}

\begin{figure}
\plottwo{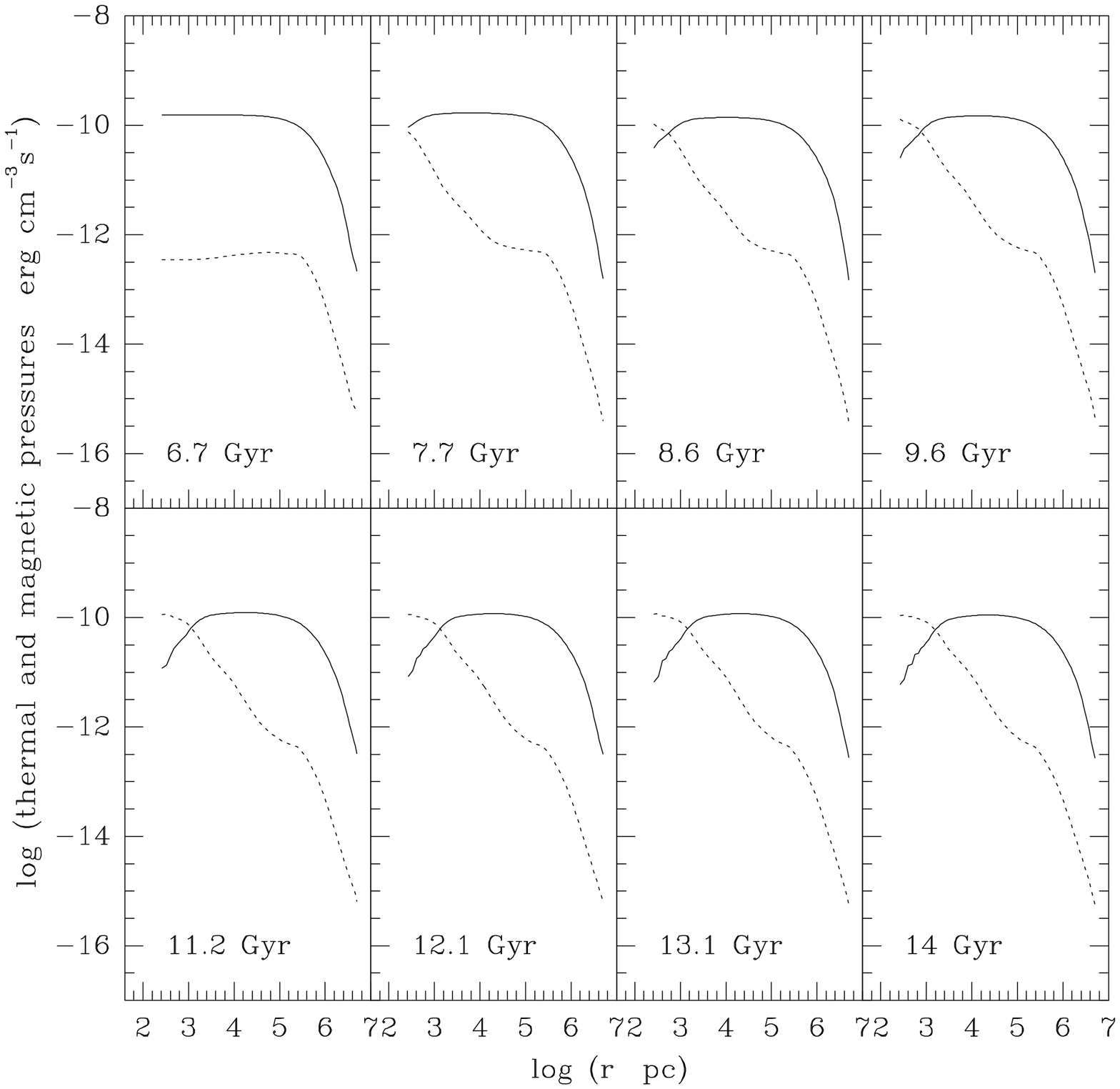}{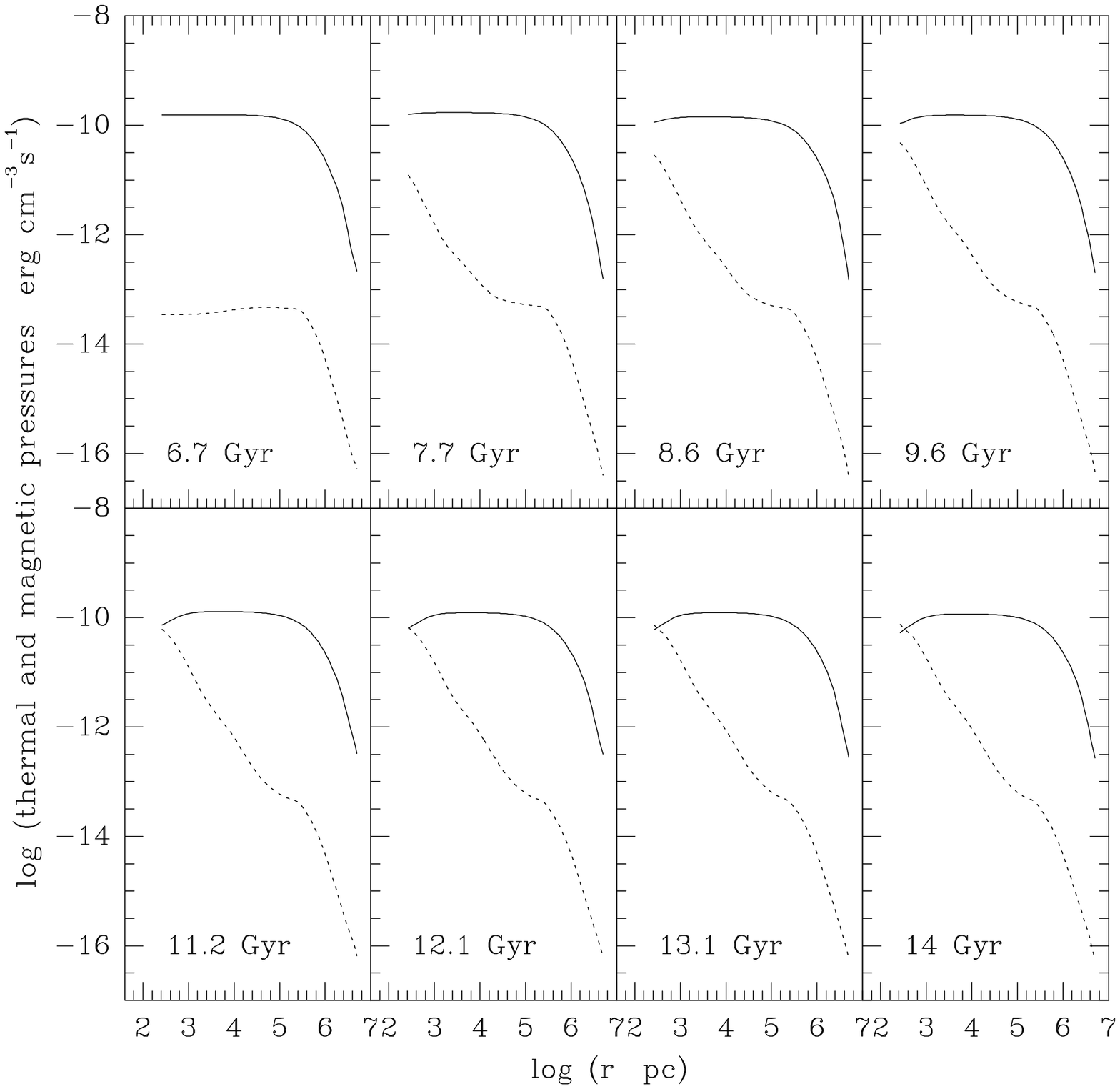}
\caption{Evolution of the magnetic (dashed line) and thermal (full line)
pressures profiles on late stages of the ICM evolution for the model A 
($\beta_0 = 10^{-2}$, left)
 and model B ($\beta_0 = 10^{-3}$, right).}
\end{figure}

Figures 1 shows the evolution of density, temperature and magnetic 
field strength, for model A, from which the presence of the cooling flow on
later stages of the ICM evolution and at inner regions, is remarkable, if
one notices the steep gradients of these quantities. 
We chose two values of
magnetic field strength for the comparison of the results, on small and 
large scales (see figure). Our results
for the magnetic field strength and also for the pressure, on large
and small scales, are in agreement to the observed ones.

Obviously the magnetic pressure (Figure 2) is compatible with the
magnetic field intensities and may be compared to the values determined by,
for instance, Bagchi et al. (1998), $p_B=B^2/8\pi \simeq 4\times 10^{-14}%
{\rm erg\;cm}^{-3}{\rm s}^{-1}$, at scales of $700\;{\rm kpc}$, in the present
time. From the analysis of the magnetic pressures expected from our models
it is clear that they agree, as well as the magnetic field strength, with
the observations. 

The present models are in many aspects similar to the one of Soker \&
Sarazin (1990). However there are two important differences between our
model and theirs: i) they take into account only small-scale magnetic field
effects; and ii) they consider the magnetic field isotropic even in the
inner regions of the cooling flow. As a matter of fact the magnetic pressure
reaches equipartition only at radius as small as $\ga 1\;{\rm kpc}$ (model A)
or $\ga 0.5\;{\rm kpc}$ (model B), because the central increase of the $\beta 
$ ratio is moderate in our model. Our more realistic description of the
field geometry is crucial. This implies that the effect of the magnetic
pressure on the total pressure of the intracluster medium, even on regions
as inner as few kpc, is small. Evolutive models for the intracluster medium, 
with a realistic calculation of the geometry and amplification of the 
magnetic fields, 
like the one presented here, indicate that magnetic pressure
does not affect the hydrostatic equilibrium, except in the innermost regions, 
i.e.  $r \la 1{\rm \;kpc}$ (see Gon\c calves \& Fria\c ca 1998 for a more 
detailed discussion).

\acknowledgements{We would like to thank the Brazilian agencies FAPESP
(97/05246-3 and 98/03639-0), CNPq and Pronex/FINEP (41.96.0908.00) 
for support.}

\end{document}